\documentclass{article}

\usepackage{arxiv}

\usepackage[utf8]{inputenc} 
\usepackage[T1]{fontenc}    
\usepackage{hyperref}       
\usepackage{url}            
\usepackage{booktabs}       
\usepackage{amsfonts}       
\usepackage{nicefrac}       
\usepackage{microtype}      
\usepackage{lipsum}
\usepackage{graphicx}
\graphicspath{ {./images/} }

\title{Exploratory Simulation of Thrombosis in a Temporary LVAD Catheter Pump within a Virtual In-vivo Left Heart Environment}

\author{
Greg W. Burgreen \\
Center for Advanced Vehicular Systems\\
Mississippi State University\\
Starkville, MS\\
\texttt{greg.burgreen@msstate.edu}\\
\And
Mansur Zhussupbekov \\
Meinig School of Biomedical Engineering\\
Cornell University\\
Ithaca, NY\\
\texttt{mz332@cornell.edu}\\
\And
Rodrigo Méndez Rojano \\
Meinig School of Biomedical Engineering\\
Cornell University\\
Ithaca, NY\\
\texttt{rm2235@cornell.edu}\\
\And
James F. Antaki \\
Meinig School of Biomedical Engineering\\
Cornell University\\
Ithaca, NY\\
\texttt{antaki@cornell.edu}\\
}

\begin{document}
\maketitle
\begin{abstract}
Percutaneous catheter pumps are intraventricular temporary mechanical circulatory support (MCS) devices that are positioned across the aortic valve into the left ventricle (LV) and provide continuous antegrade blood flow from the LV into the ascending aorta (AA). MCS devices are most often computationally evaluated as isolated devices subject to idealized steady-state blood flow conditions. In clinical practice, MCS devices operate connected to or within diseased pulsatile native hearts and are often complicated by hemocompatibility related adverse events such as stroke, bleeding, and thrombosis. Whereas aspects of the human circulation are increasingly being simulated via computational methods, the precise interplay of pulsatile LV hemodynamics with MCS pump hemocompatibility remains mostly unknown and not well characterized. Technologies are rapidly converging such that next-generation MCS devices will soon be evaluated in virtual physiological environments that increasingly mimic clinical settings. The purpose of this brief communication is to report results and lessons learned from an exploratory CFD simulation of hemodynamics and thrombosis for a catheter pump situated within a virtual \textit{in-vivo} left heart environment.
\end{abstract}


\section{Introduction}
Innovative modeling and imaging techniques are rapidly converging to improve outcomes in patients with MCS devices \cite{2023-wil}. For example, high variability in heart shapes and contractility can be characterized using cardiac computed tomography and statistical shape modeling and, when combined with CFD, can assess intracardiac flow features \cite{2022-gou}. The literature reveals a few early CFD studies involving MCS catheter pumps \cite{2001-ape,2006-tri}; one CFD study that considered an isolated catheter pump connected to the inflow of an ascending aorta \cite{2021-wan}; and only a few studies reporting any form of CFD-based blood damage modeling for a catheter pump \cite{2020-rob,2020-tia}. Blood damage in terms of hemolysis \cite{2022-zei} and thrombosis \cite{2021-sai,2023-amm} are not uncommon in catheter pumps. We have discovered no CFD publications on catheter pump performance in a clinically relevant \textit{in-vivo} setting, specifically, situated within a pulsatile LV and transporting blood into the AA. Our current efforts seek to explore and address the challenges of performing clinically relevant MCS simulations including predicting hemocompatibility-related adverse events.  

\section{Methods}
\begin{figure}[htbp!]
  \centering
  \includegraphics[width=0.30\linewidth]{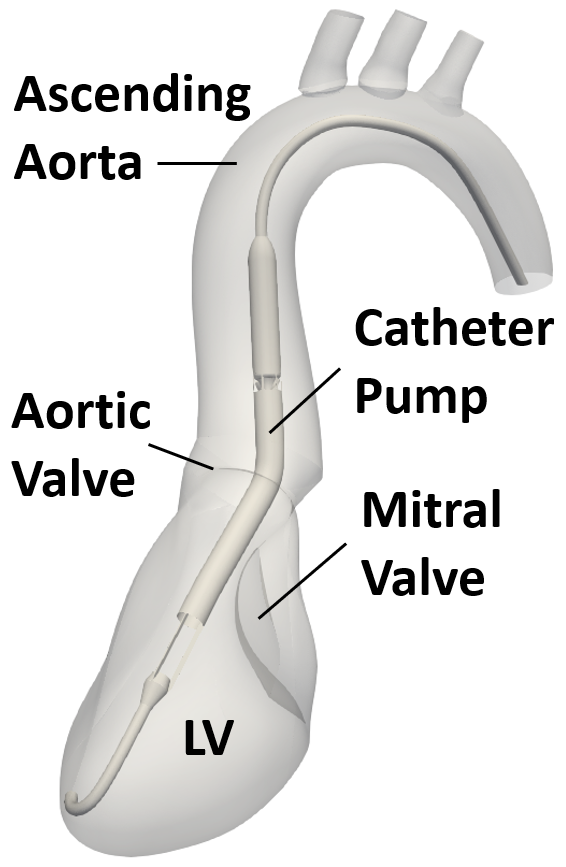}
  \caption{A catheter pump situated in a  \textit{in-vivo} environment of a non-pulsatile LV and ascending aorta.}
  \label{fig1}
\end{figure}

To this end, a generic temporary MCS catheter pump with a blood flow path inspired by the Impella 5.0 \cite{2022-zei} was situated within a static geometry consisting of an idealized LV and AA (Fig.~\ref{fig1}). We applied \textit{bloodDamageFoam}, an OpenFOAM-based CFD code with multifactorial and synergistic models of blood damage \cite{2022-bur}, to the catheter pump within this virtual \textit{in-vivo} environment to analyze its hemodynamic performance at a fixed rotor speed of 20 kRPM. The rotor had a 6.64 mm outer diameter and a 0.3 mm blade tip gap. Blood was modeled as laminar with density of 1050 kg/m3 and asymptotic viscosity of 3.6 cP. The LV and AA regions were modeled as two independent fluid domains, each with independent inflow and outlet boundary conditions. The catheter pump was modeled as a third flow domain embedded inside the LV and AA domains and connected to both domains via arbitrary mesh interfaces. The inlet cannula crossed the aortic valve boundaries and permitted blood to be pumped from the LV domain and transported into the AA domain. Boundary conditions of the LV domain consisted of steady, uniform inlet (pseudo-mitral valve) flowrate of 5.0 LPM and a constant pressure of 90 mmHg imposed at the aortic valve outlet. The AA domain had a steady inlet flowrate of 0.5 LPM at the aortic valve and constant 90 mmHg pressure at the aortic geometry outlet. The non-physiological imposition of steady flow as inlet boundary condition for the non-deforming LV domain was chosen for numerical stability reasons. An unsteady pulsatile inlet flow condition for a static LV geometry would lead to strong pump suction (i.e. negative LV pressures) and highly unrealistic intraventricular flows. The three branches of the aortic arch were modeled as capped conduits with zero exit flow. All blood-contacting surfaces were treated as non-moving walls. 

Blood damage models applied included a hemolysis model \cite{2004-kam} and a state-of-the-art thrombosis model consisting of seven biochemical and biophysical agonists that account for the effects of high shear stress-induced activation of von Willebrand Factor (vWF) \cite{2022-zhu}. All anatomical surfaces were treated as non-reactive to thrombosis, and all biomaterial surfaces of the pump system were treated as reactive to platelet deposition. The critical concentration of stretched vWF was set to 200 nmol/m3 to approximate a typical biomaterial response. 

\section{Results}

\begin{figure}[htbp!]
  \centering
  \includegraphics[width=0.90\linewidth]{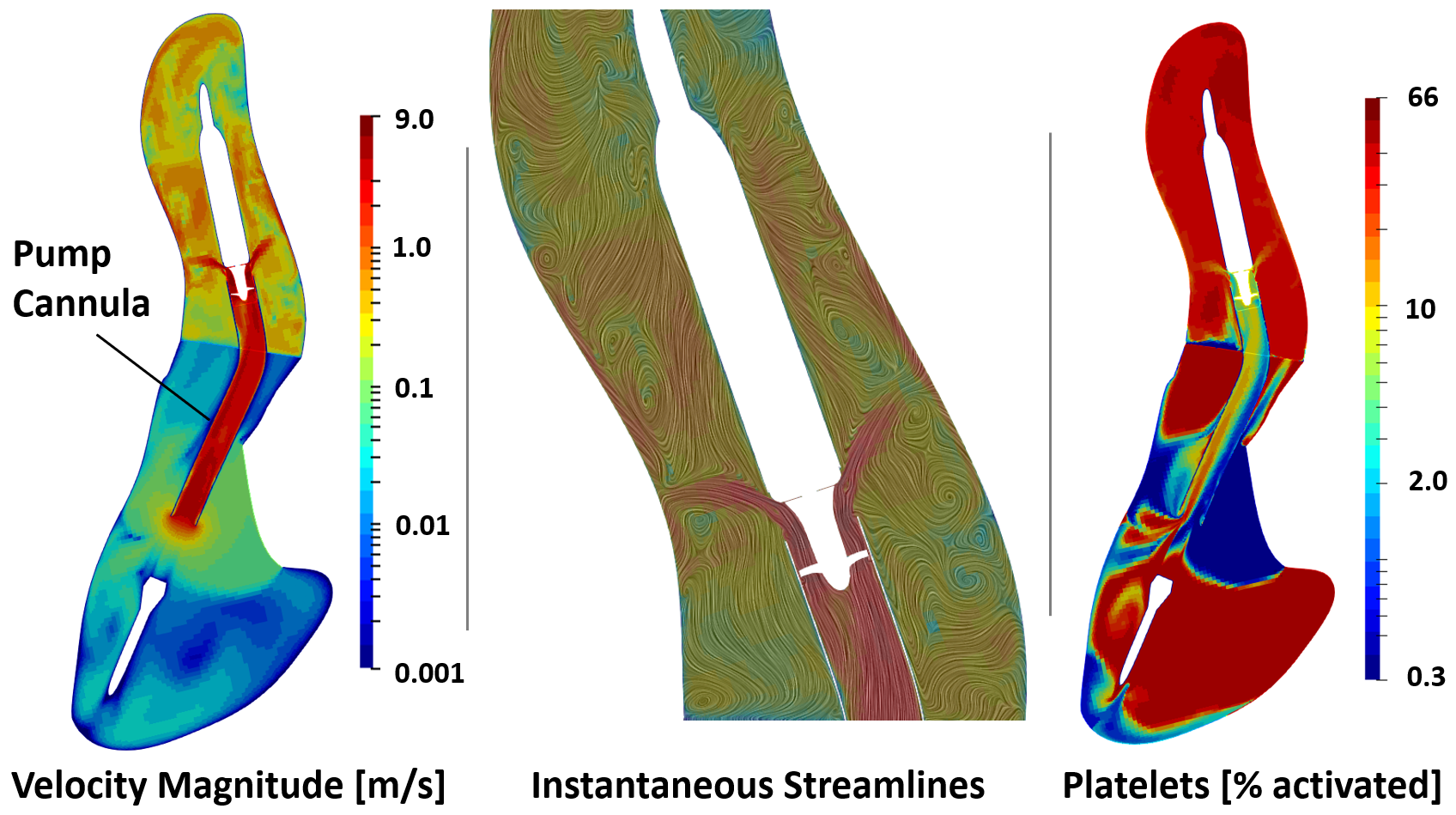}
  \caption{CFD results in coronal cut-plane passing through the centerline of the catheter pump and inlet cannula. Streamlines are colored by velocity magnitude.}
  \label{fig2}
\end{figure}

\paragraph{Hemodynamic performance.}
CFD predicted a blood flow of 4.43 LPM through the catheter pump. Thus, almost the entire steady 5.0 LPM flow into the LV was delivered into the AA by the pump, and the remaining 0.57 LPM exited the LV fluid domain via its aortic valve outlet. Low pressure (40 to 60 mmHg) was observed in the inlet cannula, and pressure throughout the entirety of the AA domain was equalized by the pump to 90 mmHg. The blood velocity field is illustrated in Figure~\ref{fig2}. The pump generated strong jets of exiting blood flow that induce multiple vortices in the ascending aorta. Blood flow relaminarized as it flowed down the descending aorta. Shear stresses were greatest in the pump rotor region, but less than 700 Pa. Blood flow within the pump cannula had high velocity and was primarily spatially uniform. 

\begin{figure}[htbp!]
  \centering
  \includegraphics[width=0.90\linewidth]{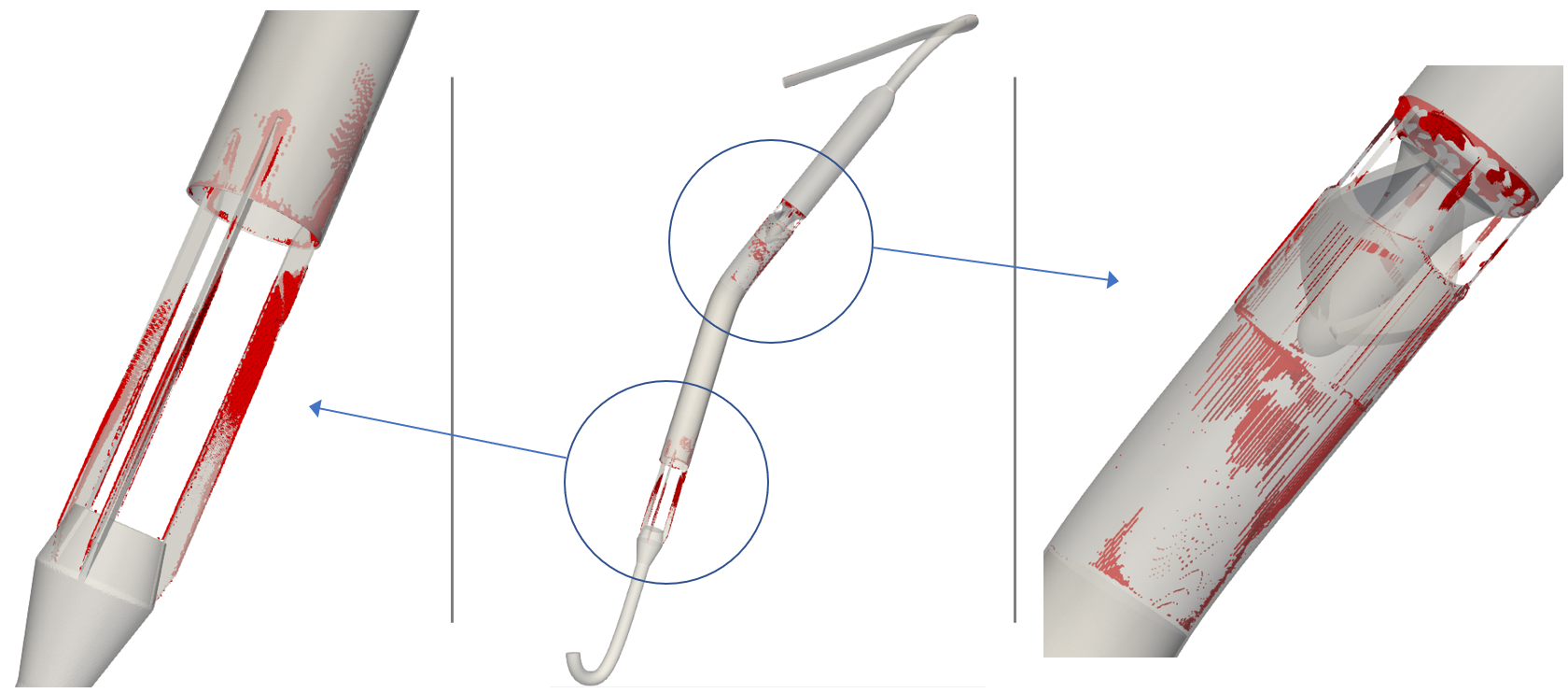}
  \caption{CFD results showing the locations of greatest thrombosis deposition.}
  \label{fig3}
\end{figure}

\paragraph{Hemocompatibility performance.}
Moderate levels of hemolysis was predicted. Due to the absence of LV contraction in this exploratory simulation, large regions of flow stasis and concomitant elevated concentrations of platelet agonists and activated platelets were observed in the lower half of the LV (Fig.~\ref{fig2}).  A mixture of resting and activated platelets was transported into the AA via the cannula. The simulation predicted 40\%-50\% of platelets in the aortic arch to be activated. The thrombosis model predicted moderate degrees of platelet deposition on the cage of the inlet cannula and within the catheter pump chamber (Fig.~\ref{fig3}). 

\section{Summary and Conclusions}

\paragraph{Lessons learned.} 
Our exploratory study had many simplifying non-physiological assumptions and yielded many lessons learned. Ongoing work is addressing several deficiencies in the present model. The first lesson learned is that the fluid domains should not be decoupled. Rather, the LV and AA should be coupled via a suitable aortic valve model that includes variable physiological patency \cite{2009-sch,2023-zin} and supports the crossing of the pump cannula. The addition of a mitral valve (MV) is needed to allow for potential pump suction-induced MV regurgitation during the isovolumetric phases of the cardiac cycle. Anatomically realistic models of LV and AA need to be considered \cite{2020-str} including dilated LVs of diseased hearts \cite{2011-bac}. Based on our exploratory study, the neglect of pulsatility and LV dynamics negatively impacts CFD predictions of  pump hemocompatibility, thus realistic LV wall motion should be modeled to produce pulsatile flow including valve dynamics \cite{2022-kar}. Options to incorporate LV pulsatility can include combinations of patient data, animal data, and simulated lumped parameter model data \cite{2022-vas}. For clinical relevance, pathological conditions such as various models of heart failure, cardiogenic shock, and valvular dysfunction should be included in advanced simulations. Such pathophysiological effects could be incorporated by coupling CFD to a model of integrative human physiology \cite{2011-hes,2022-cle}. Due to the strong fluid jet-induced mixing in the AA, fully turbulent flow simulations should be performed. Windkessel boundary conditions or lumped parameter model pressure values should be applied to all outlets of the AA to account for downstream arterial resistance. Lastly, we need to improve deficiencies of our blood damage modeling. Our current treatment of platelet deposition dynamics in rotating turbomachinery domains needs to be improved to eliminate the generation of artifactual high shear stresses within wall-attached thrombus. We need to also include the effects of fibrin-mediated platelet deposition at biomaterial surfaces in regions of low shear stress \cite{2022-men}. 

\paragraph{Conclusions.} 
This exploratory study indicated that predicted hemolysis of small high-speed catheter pumps may be minimal and that inclusion of pulsatile LV dynamics is needed to better evaluate thrombotic behaviors. This work represents our first step towards simulating hemodynamics and clinically relevant models of hemocompatibility of MCS devices within diseased pulsatile native left heart environments.

\section{Acknowledgements}
This research was supported by NIH R01 HL089456 and NIH R01 HL086918.

\section{Conflict of Interest}
No conflicts of interest are present for any of the authors. 

\bibliography{z}

\begin{thebibliography}{10}

\bibitem{2023-wil}
SI~Wilson, KE~Ingram, A~Oh, MR~Moreno, and M~Kassi.
\newblock The role of innovative modeling and imaging techniques in improving outcomes in patients with lvad.
\newblock 2023.

\bibitem{2022-gou}
L~Goubergrits, K~Vellguth, L~Obermeier, A~Schlief, L~Tautz, J~Bruening, H~Lamecker, A~Szengel, O~Nemchyna, C~Knosalla, T~Kuehne, and N~Solowjowa.
\newblock Ct-based analysis of left ventricular hemodynamics using statistical shape modeling and computational fluid dynamics.
\newblock 2022.

\bibitem{2001-ape}
J~Apel, F~Neudel, and H~Reul.
\newblock Computational fluid dynamics and experimental validation of a microaxial blood pump.
\newblock 2001.

\bibitem{2006-tri}
M~Triep, C~Brücker, W~Schröder, and T~Siess.
\newblock Computational fluid dynamics and digital particle image velocimetry study of the flow through an optimized micro-axial blood pump.
\newblock 2006.

\bibitem{2021-wan}
Y~Wang, J~Wang, J~Peng, M~Huo, Z~Yang, GA~Giridharan, Y~Luan, and K~Qin.
\newblock Effects of a short-term left ventricular assist device on hemodynamics in a heart failure patient-specific aorta model: A cfd study.
\newblock 2021.

\bibitem{2020-rob}
N~Roberts, U~Chandrasekaran, S~Das, Z~Qi, and S~Corbett.
\newblock Hemolysis associated with impella heart pump positioning: In vitro hemolysis testing and computational fluid dynamics modeling.
\newblock 2020.

\bibitem{2020-tia}
W~Tian and Y~Xiong.
\newblock Study on mechanism for preventing thrombus formation by nano-biological catheter pump.
\newblock 2020.

\bibitem{2022-zei}
R~Zein, C~Patel, A~Mercado-Alamo, T~Schreiber, and A~Kaki.
\newblock A review of the impella devices.
\newblock 2022.

\bibitem{2021-sai}
T~Saito, H~Yamamoto, S~Oishi, G~Uchino, H~Murakami, H~Kawai, and T~Takaya.
\newblock Left main trunk occlusion due to impella-related thrombus in a patient with extracorporeal cardiopulmonary resuscitation.
\newblock 2021.

\bibitem{2023-amm}
KR~Ammann, J~Ding, V~Gilman, S~Corbett, and MJ~Slepian.
\newblock Sodium bicarbonate alters protein stability and blood coagulability in a simulated impella purge gap model.
\newblock 2023.

\bibitem{2022-bur}
GW~Burgreen, R~Méndez~Rojano, M~Zhussupbekov, and JF~Antaki.
\newblock Towards a comprehensive cfd-based model of blood damage.
\newblock 2022.

\bibitem{2004-kam}
MV~Kameneva, GW~Burgreen, K~Kono, B~Repko, JF~Antaki, and M~Umezu.
\newblock Effects of turbulent stresses upon mechanical hemolysis: experimental and computational analysis.
\newblock 2004.

\bibitem{2022-zhu}
M~Zhussupbekov, R~Méndez~Rojano, WT~Wu, and JF~Antaki.
\newblock von willebrand factor unfolding mediates platelet deposition in a model of high-shear thrombosis.
\newblock 2022.

\bibitem{2009-sch}
T~Schenkel, M~Malve, M~Reik, M~Markl, B~Jung, and H~Oertel.
\newblock Mri-based cfd analysis of flow in a human left ventricle: methodology and application to a healthy heart.
\newblock 2009.

\bibitem{2023-zin}
A~Zingaro, M~Bucelli, I~Fumagalli, L~Dede', and A~Quarteroni.
\newblock Modeling isovolumetric phases in cardiac flows by an augmented resistive immersed implicit surface method.
\newblock 2023.

\bibitem{2020-str}
M~Strocchi, CM~Augustin, MAF Gsell, E~Karabelas, A~Neic, K~Gillette, O~Razeghi, AJ~Prassl, EJ~Vigmond, JM~Behar, J~Gould, B~Sidhu, CA~Rinaldi, MJ~Bishop, G~Plank, and S~Niederer.
\newblock A publicly available virtual cohort of four-chamber heart meshes for cardiac electro-mechanics simulations.
\newblock 2020.

\bibitem{2011-bac}
TN~Bachman, JK~Bhama, J~Verkaik, S~Vandenberghe, RL~Kormos, and JF~Antaki.
\newblock In-vitro evaluation of ventricular cannulation for rotodynamic cardiac assist devices.
\newblock 2011.

\bibitem{2022-kar}
E~Karabelas, S~Longobardi, J~Fuchsberger, O~Razeghi, C~Rodero, M~Strocchi, R~Rajani, G~Haase, G~Plank, and S~Niederer.
\newblock Global sensitivity analysis of four chamber heart hemodynamics using surrogate models.
\newblock 2022.

\bibitem{2022-vas}
VS~Vasudevan, K~Rajagopal, and JF~Antaki.
\newblock Application of mathematical modeling to quantify ventricular contribution following durable left ventricular assist device support.
\newblock 2022.

\bibitem{2011-hes}
RL~Hester, AJ~Brown, L~Husband, R~Iliescu, D~Pruett, R~Summers, and TG~Coleman.
\newblock Hummod: A modeling environment for the simulation of integrative human physiology.
\newblock 2011.

\bibitem{2022-cle}
JS~Clemmer and WA~Pruett.
\newblock Modeling the physiological roles of the heart and kidney in heart failure with preserved ejection fraction during baroreflex activation therapy.
\newblock 2022.

\bibitem{2022-men}
R~Méndez~Rojano, A~Lai, M~Zhussupbekov, GW~Burgreen, K~Cook, and JF~Antaki.
\newblock A fibrin enhanced thrombosis model for medical devices operating at low shear regimes or large surface areas.
\newblock 2022.

\end{thebibliography}

\bibliographystyle{unsrt}  

\end{document}